\documentclass[aps, prl, twocolumn, titlepage, showpacs]{revtex4}

\usepackage{makecell}
\usepackage{tabularx}
\usepackage{graphicx}
\usepackage{color}
\usepackage{dcolumn}
\usepackage{amsfonts}
\usepackage{amsmath}
\usepackage{bm} 
\usepackage{epstopdf}

\begin{document}
	\title{Cyclic phase transition of substrate-modulated 2D dusty plasma driven by oscillatory forces}
	
	\author{Ao Xu$^{1}$}
	\author{C.~Reichhardt$^{2}$}
	\author{C.~J.~O.~Reichhardt$^{2}$}
    \author{Yan Feng$^{1}$}
	\thanks{The author to whom correspondence may be addressed: fengyan@suda.edu.cn}
	\affiliation{$^{1}$ Institute of Plasma Physics and Technology, Jiangsu Key Laboratory of Frontier Material Physics and Devices, School of Physical Science and Technology, Soochow University, Suzhou 215006, China}
    \affiliation{$^{2}$ Theoretical Division, Los Alamos National Laboratory, Los Alamos, New Mexico 87545, USA}

\date{\today}

\begin{abstract}

Langevin dynamical simulations are performed to investigate the formation of clusters and voids of a two-dimensional-periodic-substrate (2DPS) modulated two-dimensional dusty plasma (2DDP) driven by an oscillatory force. It is discovered that, as the frequency of the oscillatory force decreases gradually, the substrate-modulated 2DDP undergoes the cyclic transition of the ordered cluster and void phases. Between the observed ordered cluster and void phases, the studied 2DDP exhibits a more uniform arrangement of particles. The discovered cyclic transition is attributed to the symmetry of the time-averaged potential landscape due to the 2DPS in the reference frame of the moving particle, as confirmed by superimposing the particle locations on the effective potential landscape under various conditions. 

\end{abstract}
		
\maketitle
	
\section{I.~Introduction}

A wide range of systems containing strongly interacting particles can be coupled to different substrates, such as charged colloids on optical traps~\cite{Reichhardt:02, Brunner:02, Mangold:03}, magnetic colloids on grooved substrates~\cite{OrtizAmbriz:16}, vortices in nanostructured type-II superconductors~\cite{Harada:96, Reichhardt:98, Grigorenko:03}, vortices in Bose-Einstein condensates with optical trap arrays~\cite{Tung:06}, cold atom systems~\cite{Bloch:05}, and various frictional systems~\cite{Reichhardt:17}. In these systems, the particle arrangement may strongly depend on the coupling strength of the particles to each other and to the substrate~\cite{Reichhardt:17}. The number of particles trapped within each potential minimum may also play an important role on their arrangement. While an additional external dc or ac driving is applied, these systems exhibit a number of interesting dynamical effects, such as depinning transition~\cite{Reichhardt:17}, soliton motion~\cite{Bohlein:12}, ordered/disordered flow phases~\cite{Gutierrez:09}, and Shapiro steps~\cite{vanLook:99}. In most studies performed to date, the number ratio of particles to substrate minima is generally not larger than five~\cite{Reichhardt:12}. As observed in the Yukawa clusters \cite{Reichhardt:12} and vortex ordering of superconductors with large blind holes~\cite{Bezryadin:96}, a large trap may include more particles inside to form a cluster with the ordered structure, while other particles located outside exhibit different properties. However, the collective behaviors of interacting particles modulated by substrates where much more particles are trapped within each potential minimum are much less well explored yet. 

When the number of particles is much greater than the number of substrate minima, the coupling between the trapped particles is fundamentally different from that for the non-trapped particles~\cite{Aranson:06, Roeller:11}. When an oscillatory driving force is applied, due to the competition of the modulation of substrate, the oscillatory driving, and the strong interaction between particles, these trapped and non-trapped particles may exhibit new collective behaviors to form different ``patterns''. It is known that, away from the equilibrium, many body systems under various forms of oscillatory driving may exhibit a wide variety of periodic pattern forming or disordered fluctuating states \cite{Aranson:06, Cross:93, Venkataramani:98, Li:03}. 

As a model system consisting of strongly interacting dust particles, dusty plasma~\cite{Chu:94, Thomas:94, Thomas:96, I:96, Melzer:96, Merlino:04, Thomas:04, Fortov:05, Feng:08, Morfill:09, Bonitz:10, Singh:22, Du:19, He:20, Beckers:23}, or complex plasma, typically refers to partially ionized gas containing micron-sized dust particles of solid matter~\cite{Feng:08}. Under the typical laboratory conditions, these dust particles are charged to $\sim -10^4 e$ ~\cite{Feng:10b} in the steady state within microseconds, as a result, the electric potential energy between neighboring dust particles is greater than their kinetic energy, i.e., they are strongly coupled, exhibiting typical properties of liquids and solids~\cite{Khrapak:21, Hartmann:14, Feng:10a, Yu:24}. Due the shielding effects of free electrons and ions, the interaction between dust particles can be accurately modeled using the Yukawa repulsion~\cite{Bonitz:06, Konopka:00, Liu:03}. These highly charged dust particles can be suspended and confined by the strong electric field in the plasma sheath, self-organizing into a single layer, i.e., the so-called two-dimensional dusty plasma (2DDP)~\cite{Feng:11, Qiao:14}. Due to the sub-millimeter level interparticle distance and the ten-Hertz level motion, the detailed trajectories of dust particles within their 2D plane can be accurately recorded using high speed cameras, and then analyzed using particle tracking~\cite{Feng:16,Feng:07,Feng:11_2}. As a result, a variety of fundamental physics procedures in liquids and solids can be studied at the kinetic level using dusty plasmas~\cite{Morfill:09}, such as transports\cite{Nosenko:04, Donko:06, Ott:09, Huang:22a, Huang:23} and phase transition~\cite{Melzer:96, Feng:08, Jaiswal:17}.

Recently, various collective behaviors of substrate-modulated 2DDP are extensively investigated using computer simulations. While a one-dimensional periodic substrate (1DPS) is applied on 2DDP, various new phenomena are observed, such as the splitting of the phonon spectra~\cite{Li:18}, structure transition~\cite{Feng:21}, and oscillation-like diffusion~\cite{Li:20}. Driven by uniform forces, the 1DPS-modulated 2DDP exhibits three dynamical states~\cite{Feng:21} of the pinned, disordered plastic flow, and moving ordered elastic, where the continuous/discontinuous transitions between them are also studied~\cite{Gu:20}. Driven by unbiased excitations, a bidirectional flow \cite{Li:23} is observed in 2DDP while the spatial symmetry of the applied 1DPS is broken. While two-dimensional periodic substrate (2DPS) modulated 2DDP is driven by uniform forces, the directional locking~\cite{Zhu:22}, superlubric-pinned transition~\cite{Huang:22}, and commensuration effect \cite{Zhu:23} are all discovered. 
However, the collective behaviors of 2DPS-modulated 2DDP driven by oscillatory forces have not been investigated yet, as we study here.

In this paper, we investigate behaviors of 2DPS-modulated 2DDP driven by oscillatory forces using computer simulations. The rest of this paper is organized as follows. 
In Sec.~II, we introduce our simulation method. In Sec.~III, we report our discovered cyclic phase transition of substrate-modulated 2DDP while monotonically decreasing the frequency of the driving oscillatory force. We also provide our interpretation of the cyclic transition, and then compare our theoretical calculations to the simulation results. In Sec.~IV, we provide a brief summary of our discovery reported here.

\section{II.~Simulation method}
	
We perform Langevin dynamical simulations~\cite{Li:18}
to mimic a substrate-modulated 2DDP driven by oscillatory forces. For a particle $i$, the equation of motion is
\begin{equation}
	\label{Langevin}
	m\ddot{\textbf{r}}_i=-\sum \bm{\nabla} \phi _{ij}-\nu m\dot{\textbf{r}}_i+\xi _i(t)+\textbf{F}_i^{\rm S}+\textbf{F}_i^{\rm D}.
\end{equation}
Here, $\phi _{ij}=Q^2$exp$(-r_{ij}/\lambda _{\rm D})/4\pi \epsilon _0r_{ij}$ is the interparticle Yukawa repulsion~\cite{Kalman:04}, $Q$ is the particle charge, $\lambda _{\rm D}$ is the Debye length, and $r_{ij}$ is the distance between particles $i$ and $j$. The second term $\nu m\dot{\textbf{r}}_i$ on the right hand side of Eq.~(1)
is the frictional gas drag~\cite{Liu:03a}, which is proportional to the particle momentum. The third term $\xi _i(t)$ represents the Langevin random kicks~\cite{Yu:22}, assumed to have a zero-mean Gaussian distribution of magnitude given by the fluctuation-dissipation theorem~\cite{Gunsteren:82}. 

The fourth term $\textbf{F}_i^{\rm S}$ is the force from the 2DPS, which consists of $N_p$ non-overlapping potential wells placed in a triangular array. The substrate potential is $\psi(\textbf{r})=\Sigma_{k=1}^{N_p} \psi_k$, where $\psi_k =1/2(F_p/r_p)(\textbf{r}-\textbf{r} _k^{(p)})^2\Theta (r_p-|\textbf{r}-\textbf{r} _k^{(p)}|)$~\cite{Reichhardt:06}, so that $\textbf{F}_i^{\rm S}=-\bm{\nabla} \psi(\bf{r})$.
Here, $\textbf{r} _k^{(p)}$ is the location of the center of the $k$th potential well, $r_p$ is the radius of each well, $F_p$ corresponds to the magnitude of the force, and $\Theta$ is the Heaviside step function. The last term in Eq.~(1) is the applied oscillatory driving force with the expression of $\textbf{F}_i^{\rm D}=A[\sin (\omega t)\hat{\textbf{x}}+\cos (\omega t)\hat{\textbf{y}}]$, where $A$ is the amplitude and $\omega$ is the angular frequency.

Our simulated 2DDP is characterized by two dimensionless parameters~\cite{Morfill:09, Bonitz:10, Kalman:04}, the coupling parameter
$\Gamma =Q^2/(4\pi \epsilon _0ak_{\rm B}T)$
and the screening parameter 
$\kappa =a/\lambda _{\rm D}$.
Here $T$ is the kinetic temperature of particles and $a=(n\pi)^{-1/2}$~\cite{Kalman:04} is the Wigner-Seitz radius for the areal number density $n$. In our simulations, we always specify $\Gamma=300$ and $\kappa=2$, corresponding to a typical liquid state~\cite{Hartmann:05, Yu:24} in the absence of substrates or driving. Here, the length and time scales are normalized to be dimensionless using $a$ and the inverse of the nominal 2DDP frequency $\omega _{{\rm pd}}^{-1}=(Q^2/2\pi \epsilon _0ma^3)^{-1/2}$~\cite{Kalman:04}, respectively, while forces are normalized using $F_0=Q^2/(4\pi \epsilon _0a^2)$. 

Here are the details of our simulations. We place $N = 1024$ particles in a rectangular box with the size of $61.1a \times 52.9a$, using the periodic boundary conditions in the both directions. The 2DPS consists of $N_p=36$ potential wells with the radius $r_p=4a$ and maximum depth of $ 2aF_0$. The magnitude of the applied oscillatory force is specified as $A/F_0 = 20.0$, and the frequency is varied over the range from $0.1$ to $4.0~\omega _{{\rm pd}}$ with a step of $0.05~\omega _{{\rm pd}}$. The frictional damping coefficient is set to $\nu =0.027\omega _{{\rm pd}}$, which is typical in 2D dusty plasma experiments~\cite{Liu:03}. We set the cutoff radius of the Yukawa interaction to $24.8a$. For each simulation run, we first integrate Eq.~(1) for more than $10^5$ steps with a time step of $0.002\omega _{{\rm pd}}^{-1}$~\cite{Liu:05} to reach a steady state. Then, we record the particle positions and velocities during the next two hundred periods of the oscillatory force for the data analysis reported here. Other simulation details are the same as in~\cite{Huang:22,Zhu:22}. In addition to the simulations described above, we also performed a few test runs with different numbers of particles and potential wells to confirm that our reported results are independent of the system size.

\section{III.~Results}
\begin{figure}[htp]
\centering
\includegraphics[width=85mm]{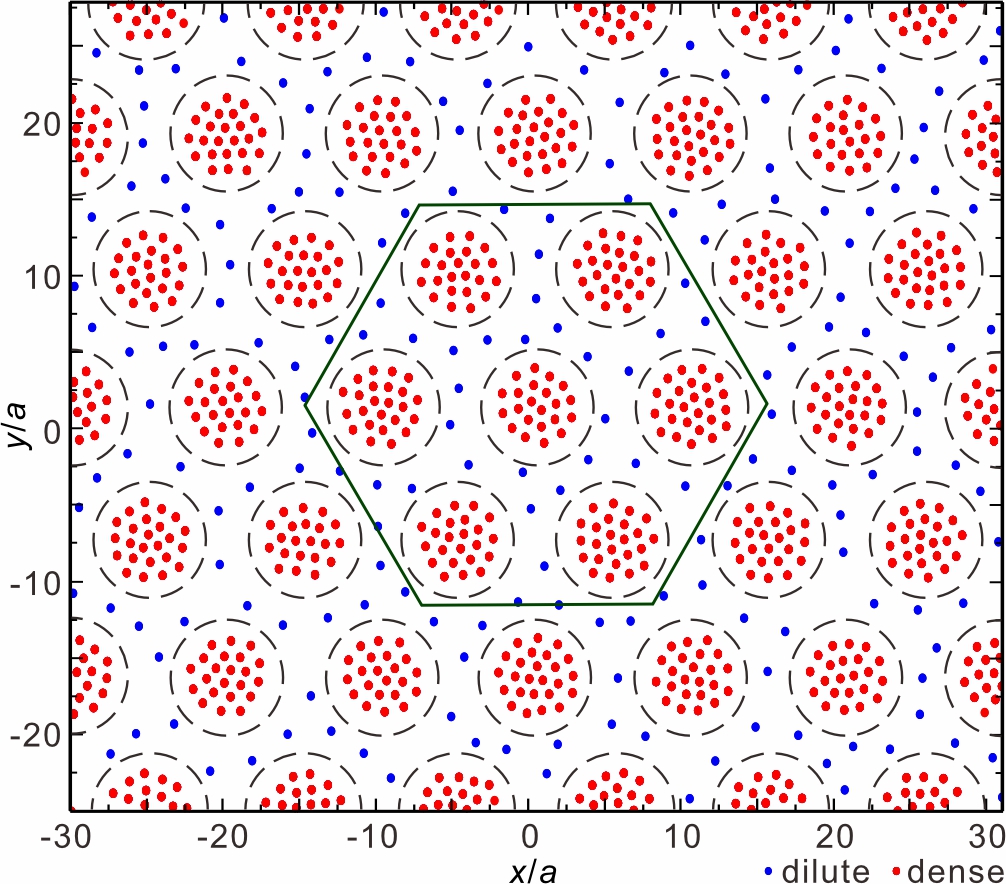}
\caption{Snapshots of the particle positions (dots) of the substrate-modulated 2DDP without a driving force. Inside potential wells (dashed circles) of the substrate, particles are mostly caged, forming a dense cluster, which we term ``dense'' particles. While the remaining particles are sparsely scattered in the interstitial region between potential wells, termed as ``dilute'' particles here. The seven wells enclosed by the hexagon will be used in the latter analysis in Fig.~\ref{figure4}. Note, we distinguish between dilute and dense particles using Eq.~(2). The simulation conditions are $\Gamma =300$, $\kappa =2$, $F_p/F_0=1.0$, $r_p/a=4$, and $N_p=36$.} 
\label{figure1}
\end{figure}
 
\subsection{A. Structure of 2DDP on 2DPS}

We first investigate the arrangement of particles modulated by the applied 2DPS without an oscillatory force. Figure~\ref{figure1} shows a typical snapshot of the instantaneous particle arrangement along with the locations and effective ranges of the potential wells of the 2DPS. From Fig.~\ref{figure1}, the particles are clearly divided into two groups, one stays inside potential wells of the 2DPS, while the other sits between potential wells. Inside each potential well, there are dozens of captured particles, forming a dense cluster, which we term ``dense'' particles here. The remaining particles are sparsely scattered in the interstitial region between these potential wells, which we term ``dilute'' particles. From our observation, the total number of dilute particles almost never changes, clearly indicating that these dilute particles do not enter potential wells for our studied conditions. 

To quantitatively distinguish these two types of particles, we employ the concept of ``solid concentration''~\cite{Wang:08} to introduce a threshold length $L_{\rm lower}$, defined as
\begin{equation}
L_{\rm lower} = \langle\overline{L}\rangle-c \langle \sigma (L)\rangle.
\end{equation}
Here, $\overline{L}$ represents the averaged length of the nearest-neighboring bonds in the liquid 2DDP of $\Gamma=300$ and $\kappa=2$ without a 2DPS or an oscillatory force, while $\sigma (L)$ is the standard deviation of the lengths of the nearest-neighboring bonds for all particles from our simulation. Note, the notation $\langle \rangle$ indicates an ensemble average. Next, for each particle, we compare the average of its two shortest bonds $\overline{L_{i}}$ to the threshold length $L_{\rm lower}$. If $\overline{L_{i}} < L_{\rm lower}$, then this particle is identified as a dense particle; otherwise, it is marked as a dilute particle. From~\cite{Wang:08}, the value of the dimensionless constant $c$ is chosen empirically for different systems. Here, we find that the value $c = 1.75$ is able to correctly distinguish the dense and dilute particles, as shown in Fig.~\ref{figure1}.

\subsection{B. Substrate-modulated 2DDP driven by oscillatory forces}

\begin{figure}[htp]
\centering
\includegraphics[width=85mm]{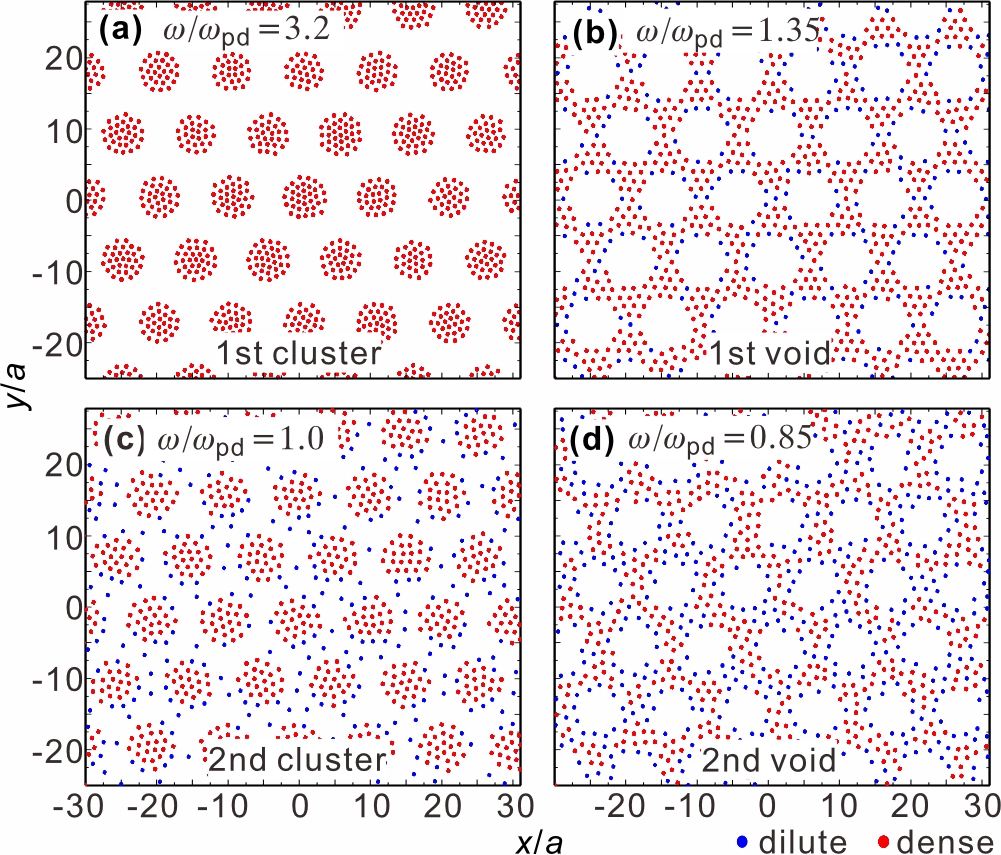}
\caption{Typical snapshots of particle positions of the same substrate-modulated 2DDP as in Fig.~1 driven by an oscillatory force of $A/F_0=20.0$ with the varying frequency of $\omega/\omega _{{\rm pd}}=3.2$ (a) to 1.35 (b), 1.0 (c), and 0.85 (d).
When $\omega/\omega _{{\rm pd}}=3.2$ in panel (a), the particles accumulate to form the ordered triangular-arranged clusters, termed as ``1st cluster.'' When $\omega/\omega _{{\rm pd}}=1.35$ in panel (b), the particles are distributed to form the ordered triangular-arranged voids, termed as ``1st void.'' When $\omega/\omega _{{\rm pd}}=1.0$ in panel (c), the ordered triangular-arranged clusters are formed again, with a few dilute particles among clusters, termed as ``2nd cluster.'' When $\omega/\omega _{{\rm pd}}=0.85$ in panel (d), the ordered triangular-arranged voids are formed again with more dilute particles, termed as ``2nd void.'' }
\label{figure2}
\end{figure}

As the major result of this paper, we discover a cyclic transition between the cluster and void phases of substrate-modulated 2DDP, while the driving frequency decreases monotonically. This discovery is illustrated in Fig.~\ref{figure2}, where we present the instantaneous particle arrangements of the substrate-modulated 2DDP driven by an oscillatory force with a constant amplitude of $A/F_0 = 20.0$ while the frequency $\omega$ varies. For $\omega=3.2\omega _{{\rm pd}}$ in Fig.~\ref{figure2}(a), all of the particles accumulate to form distinct ordered triangular-arranged clusters, which we term the ``cluster phase''. As the frequency of the oscillatory force decreases, the boundaries of these clusters gradually blur until the clusters completely disappear, so that a uniform arrangement of a liquid state emerges. When the frequency decreases further, voids gradually appear in the particle arrangement. When $\omega=1.35 \omega_{{\rm pd}}$, in the particle arrangement, quite a few voids with distinctive edges are arranged in a triangular structure with the hexagonal symmetry, as presented in Fig.~\ref{figure2}(b), which we term the ``void phase''. As the frequency decreases further, the cluster and void phases each appear a second time, as shown in Fig.~\ref{figure2}(c) and (d), while the edges of the clusters and voids are more blurred, as compared with the phases at higher driving frequencies.

Due to the same applied oscillatory force, from our observation, all particles in these cluster and void phases in Fig.~\ref{figure2} gyrate as a single object in the steady state, with nearly negligible relative motion between individual particles and clusters or voids. As the frequency of the oscillatory force decreases from $\omega/\omega _{{\rm pd}} = 4.0$ to 1.0, the gyroscopic radius of each individual particle increases monotonically. When the gyroscopic radius is small, as in Fig.~\ref{figure2}(a), all of the particles remain confined inside the potential wells, so that each cluster occupies one potential well of the 2DPS. For the lower frequency cluster phase in Fig.~\ref{figure2}(c), the gyroscopic radius is much larger, so that all of the particles are able to move both inside and outside of the potential wells; however, most of the particles still tend to aggregate together to form clusters. From our observation, for the conditions with voids as in Figs.~\ref{figure2}(b) and (d), the exhibited voids never overlap with the potential wells at all. From our understanding, during the gyroscopic motion of particles with different magnitudes, the combination effect of their interaction and the applied 2DPS results in the clusters and voids in Fig.~\ref{figure2}, as well as the uniform arrangement of particles.

\begin{figure}[htp]
\centering
\includegraphics[width=85mm]{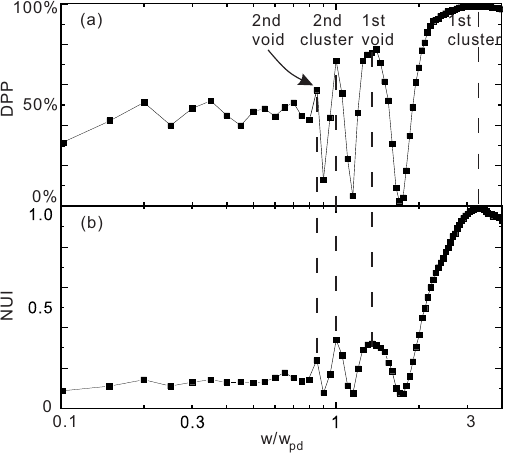}
\caption{Variations of our calculated dense particle proportion (DPP) (a) and non-uniformity index (NUI) (b) of substrate-modulated 2DDP with the frequency of the oscillatory force. Clearly, DPP and NUI vary synchronously, so that the frequencies corresponding to their maxima and minima are the same. Their four peaks correspond to the cluster and void phases, as marked there. Note that the amplitude of the oscillatory force is always specified as $A/F_0=20.0$.
} 
\label{figure3}
\end{figure}

To quantitatively characterize the cyclic transition between the cluster and void phases, we introduce two diagnostics, the dense particle proportion (DPP) and the non-uniformity index~\cite{Zhu:01} (NUI). The DPP is defined as
\begin{equation}
    \textbf{DPP}=\langle N_{\rm dense}\rangle/N.
\end{equation}
Here, $N_{\rm dense}$ is the number of dense particles determined using Eq.~(2), while $N=1024$ is the total number of particles in our simulations. Note that $\langle \rangle$ represents an ensemble average. From Eq.~(3), a higher DPP value means a higher proportion of dense particles, corresponding to more significant aggregation of particles. In fact, this aggregation is present in both the cluster and void phases from Fig.~\ref{figure2}, so that the related DPP values should be higher. In contrast, for uniformly distributed particles with no aggregation, the corresponding DPP value should be nearly zero.

Our obtained DPP results with the varying frequency $\omega / \omega _{{\rm pd}}$ of the oscillatory force are presented in Fig.~\ref{figure3}(a). When $3.2 \leq \omega/ \omega _{{\rm pd}} \leq 4.0$, the DPP value $\approx 100\%$, indicating nearly complete aggregation of the particles at high driving frequencies. The maximum value of DPP occurs at $\omega/ \omega _{{\rm pd}}=3.2$, corresponding to the 1st cluster phase illustrated in Fig.~\ref{figure2}(a), also labeled as the $1$st cluster in Fig.~\ref{figure3}(a). When $0.85 < \omega/ \omega _{{\rm pd}} < 3.2$, the DPP results exhibit several remarkable peaks at the frequencies of $\omega / \omega _{{\rm pd}} \approx $ 1.35, 1, and 0.85, respectively. From the corresponding particle distributions at these frequencies, we confirm that the void and cluster phases occur repeatedly, labeled as $1$st void, $2$nd cluster, and $2$nd void in Fig.~\ref{figure3}(a). All of the four labeled phases
in Fig.~\ref{figure3}(a) are illustrated in Fig.~\ref{figure2}. In the frequency range of $0.85 < \omega/ \omega _{{\rm pd}} < 3.2$, the DPP results also exhibit three valleys with significantly low values, even close to zero, corresponding to the pretty uniform arrangement of particles.

To characterize the particle arrangement, in Fig.~\ref{figure3}(b), we use the second diagnostic of NUI, which is defined as
\begin{equation}
    \textbf{NUI}=\langle\sigma (n(x,y))\rangle/\sigma _{max}(n(x,y)).
\end{equation}
Here, $n(x,y)$ is the areal number density of the 2DDP as a function of coordinates $x$ and $y$ for each snapshot, while $\sigma (n(x,y))$ is the standard deviation of these $n(x,y)$ values. Note, to calculate the instantaneous value of $n(x,y)$, we divide the system into $20 \times 20$ cells, resulting in 400 values of the areal number density. Following the procedure in~\cite{Zhu:01}, we choose a normalization factor $\sigma _{max}(n(x,y))$ of $0.137$, which is the maximum value reached by $\sigma (n(x,y))$ across all frequencies of the driving oscillatory force in the range of $0.1 \leq \omega/\omega _{{\rm pd}} \leq 4.0$. To obtain the ensemble average $\langle \rangle$ in this case, we use the time average.

Figure~\ref{figure3}(b) shows that the four large peaks in NUI located at $\omega / \omega _{{\rm pd}}=$ 3.2, 1.35, 1, and 0.85 perfectly match the four DPP peaks, as indicated by the dashed lines. When our studied 2DDP exhibits significant cluster and void phases, the strong non-uniformity in the particle arrangement reasonably leads to higher values of NUI, from its definition of Eq.~(4). As shown in Fig.~\ref{figure3}(b), the three valleys of the NUI results occur at the same frequency values as the minima of the DPP, clearly indicating the uniform arrangement of particles there.

\subsection{C. Phase transition interpretation}

To interpret our discovered cyclic phase transition of the substrate-modulated 2DDP driven by oscillatory forces, we study the collective dynamics of particles in the reference frame of the moving particle. From our simulation results, all of the particles undergo gyroscopic motion nearly synchronously, as in a rigid body, due to the fact that the oscillatory driving force $\text{F}_i^D$ is about 70 times larger than the repulsion between neighboring particles or the confining force of the 2DPS. Due to this profound synchronized gyroscopic motion, in the traditional rest reference frame, none of the particle dynamics modulated by the 2DPS, the stochastic thermal motion of particles, and their diffusive motion can be easily identified any more. Thus, to simplify the interpretation, we shift to the reference frame of the moving particle to study the collective dynamics. In this frame, our applied 2DPS is no longer stationary, but instead undergoes the gyroscopic motion in the direction opposite to the motion of the particles. This operation allows the particle dynamics modulated by the 2DPS, the stochastic thermal motion of particles, and their diffusive motion to be identified easily. Note, the moving particle reference frame does not rotate, as a result, in this frame the inertial force acting on particles is just the inverse of the oscillatory force.

\begin{figure}[htp]
\centering
\includegraphics[width=85mm]{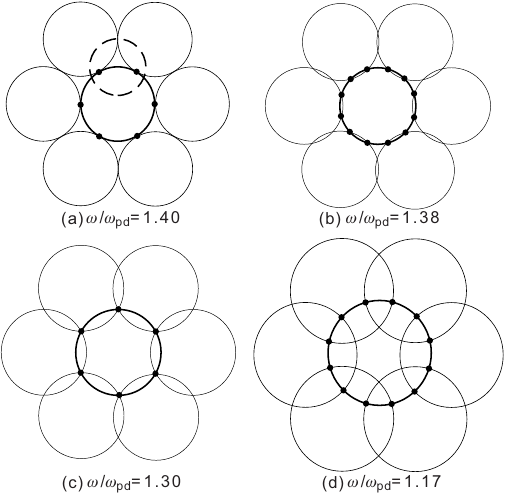}
\caption{
Sketches of orbits (solid circles) of the centers of 7 nearest-neighboring potential wells of the 2DPS viewed in the moving particle reference frame for the driving frequencies of $\omega / \omega _{{\rm pd}}=1.40$ (a), $1.38$ (b), $1.30$ (c), and $1.17$ (d). In panels (a) and (c), the orbits of the six nearest neighbors evenly divide the central orbit into 6 parts, leading to the hexagonal symmetry. In panels (b) and (d), the central orbit is evenly divided into 12 parts, resulting in the dodecagonal symmetry. Note, in panel (a), we draw the central potential well inside the hexagon in Fig.~\ref{figure1} at one instant using a dashed circle.
}
\label{figure4}
\end{figure}

In the moving particle reference frame, the modulation of the applied 2DPS under the oscillatory force with different frequencies exhibits different symmetries in the potential landscape, as shown in Fig.~\ref{figure4}. A single particle that experiences only the oscillatory force would perform the gyroscopic motion with a radius of $R=A/m\omega^2$. As a result, in the moving particle reference frame, the 2DPS undergoes the gyroscopic motion with the radius of $R=A/m\omega^2$ in the opposite direction. Clearly, as the frequency $\omega$ decreases, the radius of the gyroscopic motion increases monotonically, resulting in different symmetries of the intersections of the gyroscopic orbits. In Fig.~\ref{figure4}, we present the orbits of seven potential well centers viewed in the moving particle reference frame for the driving frequencies of $\omega/ \omega _{{\rm pd}}=$ (a) 1.40, (b) 1.38, (c) 1.30, and (d) 1.17, respectively. When $\omega / \omega _{{\rm pd}}=$ 1.40 and 1.30, as shown in Figs.~\ref{figure4}(a) and (c), the orbits of the six nearest neighbors evenly divide the central orbit into six parts, producing the hexagonal symmetry. When $\omega / \omega _{{\rm pd}}=$ 1.38 and 1.17, as shown in Figs.~\ref{figure4}(b) and (d), the central orbit is evenly divided into twelve parts, resulting in the dodecagonal symmetry. 

\begin{table}[htp]
    \centering
    \tabcolsep=3.4mm
    \setcellgapes{7pt}
    \makegapedcells
    \begin{tabular}{c|c c| c c| c}
    \hline
    \hline
    \multicolumn{6}{c}{\makecell{symmetry of intersection distribution}}\\
    \hline
    \multicolumn{1}{c}{\makecell{}} & & \multicolumn{2}{c}{\makecell{hexagonal}} & \multicolumn{2}{c}{\makecell{dodecagonal}}\\
    \hline
    $\beta$ &  & \multicolumn{1}{c}{\makecell{$\theta = 0$}} & $\pi/3$ & \multicolumn{1}{c}{\makecell{$\pi/6$}} & $\pi/2$\\  \hline
      & 1st cluster & $>3.03$&    &   \multicolumn{1}{c}{\makecell{}}   &     \\  \cline{2-5}
    1 &   1st void  & \multicolumn{1}{c}{\makecell{1.40}} & 1.30 & 1.38 & 1.17\\  \cline{2-5}
    2 & 2nd cluster & \multicolumn{1}{c}{\makecell{1.06}} & 1.0  & 1.05 & 0.89\\ 
    3 &             & \multicolumn{1}{c}{\makecell{0.99}} & 0.93 & 0.98 & 0.83\\  \cline{2-5}
    4 &  2nd void   & 0.86 & ...  & \multicolumn{1}{c}{\makecell{...}}  & ... \\  
    \hline
    \hline
    \end{tabular}
    \caption{Frequencies $\omega / \omega _{{\rm pd}}$ derived from Eq.~(5) of the driving oscillatory force at which the hexagonal and dodecagonal symmetries are expected to arise in the intersections of the orbits of the potential well centers in the moving particle reference frame.}
    \label{table 1}
\end{table}

As the frequency of the oscillatory force further decreases, the gyroscopic radius of the potential wells increases. Therefore, for the hexagonal and dodecagonal symmetries studied here, we must also consider contributions from the second- and third-nearest neighbors. From our derivations, as provided in Appendix A, we obtain an expression of the gyroscopic radius corresponding to the hexagonal and dodecagonal symmetries as
\begin{equation}
    \frac{A}{m\omega^2 d}=\frac{\sqrt{\beta^2+2\beta+2\times2^{(-1)^\beta}}}{4\cos(\theta/2)}.
\end{equation}
Here, $d=10.183a$ is the distance between the centers of neighboring potential wells in the triangular array of our applied 2DPS. The number $\beta$ corresponds to the $\beta$th-nearest neighbors. The value of $\theta$ has only two choices, i.e., $\theta=0$ and $\pi/3$ for the hexagonal symmetry, or $\pi/6$ and $\pi/2$ for the dodecagonal symmetry, as described in detail in Appendix A. Note, our findings in Fig.~\ref{figure4} perfectly match the results of Eq.~(5) for $\beta = 1$.

Following Eq.~(5), we obtain the values of $\omega/\omega _{{\rm pd}}$ in Table~I for different symmetries of the intersections on the orbits of potential well centers. Here, we substitute $\beta$ values of 1, 2, 3, and 4 into Eq.~(5), for four different values of $\theta$ corresponding to the hexagonal and dodecagonal symmetries, respectively. The only exception is $\omega/ \omega _{{\rm pd}} > 3.03$, which we derive from the conditions under which the orbits of the potential well centers in the moving particle reference frame never overlap. Since the values of $\omega/ \omega _{{\rm pd}}$ with different symmetries are determined geometrically in Table~I, we would like to verify the corresponding particle distribution from our simulations. A comparison of Table~I to our DPP and NUI results indicates that the cluster and void phases shown in Fig.~\ref{figure3} match either the hexagonal symmetry for $\theta = 0$ and $\pi/3$ as in Fig.~\ref{figure4}(a) and (c), or the dodecagonal symmetry only for $\theta = \pi/6$ as in Fig.~\ref{figure4}(b), as labeled in Table~I. However, the dodecagonal symmetry for $\theta = \pi/2$ as in Fig.~\ref{figure4}(d) corresponds to the valley of the DPP or NUI results in Fig.~\ref{figure3}, where the particle arrangement is pretty uniform.

\begin{figure*}[htp]
    \centering
	\includegraphics[width=165mm]{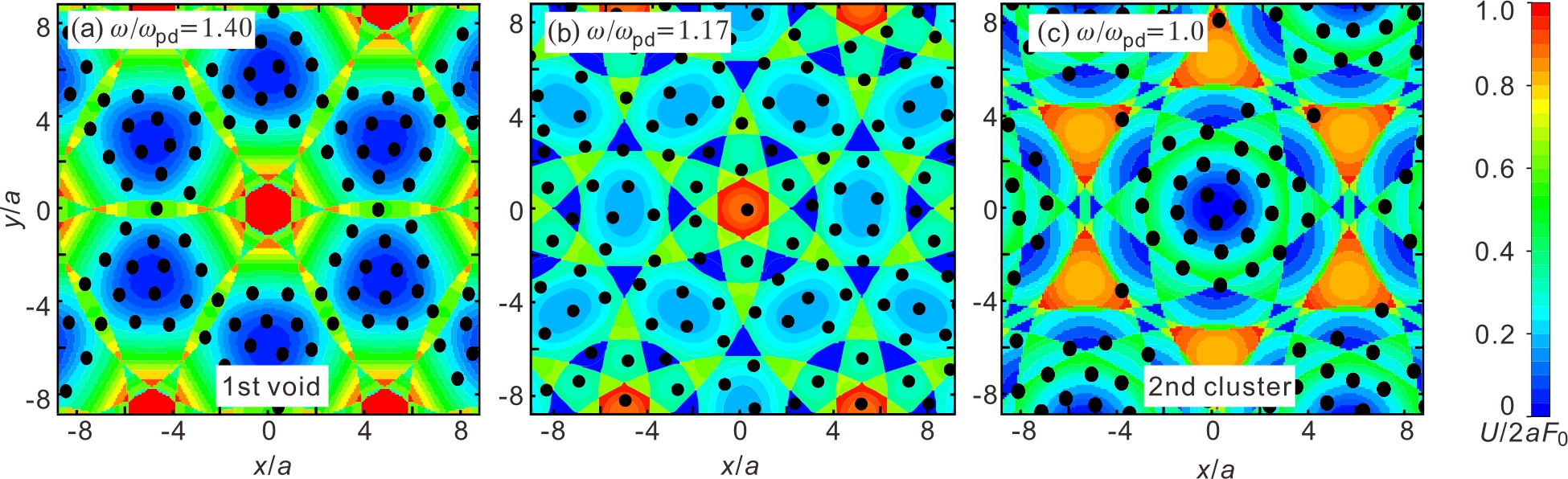}
\caption{Contour plot of the time-averaged potential landscape in the reference frame of the moving particle superimposed by the particle locations (dots), for the different frequencies of the driving oscillatory force $\omega / \omega _{{\rm pd}}=1.40$ (a), $1.17$ (b), and $1.0$ (c). In panels (a) and (c), the hexagonal symmetry of the effective substrate almost perfectly matches the arrangement of clusters and voids from the particle arrangement. In panel (b), the dodecagonal symmetry of the effective substrate corresponds to many more minima of potential wells with smaller size, reasonably leading to a more uniform arrangement of particles.}
\label{figure5}
\end{figure*}

To further confirm our interpretation, in Fig.~\ref{figure5}, we superimpose the particle arrangements above the time-averaged potential landscapes in the reference frame of the moving particle. We plot the potential landscape for the hexagonal symmetry of $\omega / \omega _{{\rm pd}}=1.40$ in Fig.~\ref{figure5}(a), corresponding to the 1st void in Fig.~\ref{figure3}. From Fig.~\ref{figure5}(a), six potential minima form a regular hexagon, whose center acts as a potential barrier. As a result, the particles tend to aggregate near the effective potential minima, resulting in the formation of a void in the center. In Fig.~\ref{figure5}(c), we plot the potential landscape for the hexagonal symmetry of $\omega / \omega _{{\rm pd}}=1.0$, corresponding to the 2nd cluster in Fig.~\ref{figure3}. From Fig.~\ref{figure5}(c), six potential barriers form a regular hexagon with a potential minimum in the center. Thus, the particles tend to be confined by the central potential minimum, forming a cluster. Only a few particles scatter outside of the potential minimum. We plot the potential landscape for the dodecagonal symmetry at $\omega / \omega _{{\rm pd}}=1.17$ in Fig.~\ref{figure5}(b). Compared with Figs.~\ref{figure5} (a) and (c), the number of effective potential minima in Fig.~\ref{figure5}(b) is much larger. Each is of such small size, so that it is almost impossible to confine more than two particles within one effective potential minimum in Fig.~\ref{figure5}(b), reasonably leading to the nearly uniform arrangement of particles. Note, we also confirm that, under other conditions not presented here, the potential landscape has the similar hexagonal symmetry for the cluster and void phases, or exhibits similar smaller effective potential minima for the pretty uniform particle arrangement.

Our interpretation indicates that the formation of our observed cluster/void phases or nearly uniform arrangement is determined by the time-averaged potential landscape viewed from the reference frame of the moving particle. As the frequency of the oscillatory force $\omega$ varies, the geometry of the potential landscape changes simultaneously following Eq.~(5), as also presented in Table~I. For the values of $\omega$ corresponding to $\theta = 0$, $\pi/6$, and $\pi/3$, the potential landscape has the hexagonal symmetry, resulting in the ordered cluster and void phases, as in Figs.~\ref{figure5}(a) and (c). For the values of $\omega$ corresponding to $\theta = \pi/2$, the potential landscape contains much more effective potential minima with smaller sizes, almost impossible for particles to aggregate, resulting in a more uniform arrangement as in Fig.~\ref{figure5}(b). 

\section{IV.~Summary}
In summary, we perform Langevin dynamical simulations to investigate collective behaviors of substrate-modulated 2DDP driven by oscillatory forces. In the absence of the oscillatory force, particles are divided into two groups, which are dense and dilute particles. Dense particles stay inside potential wells to form ordered triangular-arranged clusters, while dilute particles are sparsely scattered in the interstitial region between the potential wells. When the oscillatory force is applied on the substrate-modulated 2DDP, we discover the cyclic transition between the cluster and void phases. When the frequency of the oscillatory force decreases from $4.0\omega _{{\rm pd}}$ to $0.1\omega _{{\rm pd}}$ monotonically, we find the substrate-modulated 2DDP undergoes the transition from the ordered cluster, to the uniform distribution, next to the void, then to the cluster-uniform-void cycle again, i.e., repeating in this phase transition cycle. 

We attribute our observed cyclic transition of these ordered cluster and void phases to the symmetry of the time-averaged potential landscape viewed from the reference frame of the moving particle. The symmetry of this time-averaged potential landscape is dependent on the frequency of the applied oscillatory force, as derived in Eq.~(5) and also presented in Table~I. By superimposing the particle positions from our simulations above the plotted time-averaged potential landscape in Fig.~\ref{figure5}, we confirm that our observed cyclic transition between the cluster and void phases can be interpreted by the effective potential landscape viewed from the reference frame of the moving particle. Beyond dusty plasmas, for other particle systems where particles are driven by a uniform ac force, our treatment from the reference frame of the moving particle probably great simplifies the investigation of the complicated self-organization of the system, just like our observed cyclic transition of the ordered cluster and void phases here.

\section*{ACKNOWLEDGMENTS}
The work was supported by the National Natural Science Foundation of China under Grant No. 12175159, the 1000 Youth Talents Plan, the Priority Academic Program Development of Jiangsu Higher Education Institutions, and the U. S. Department of Energy through the Los Alamos National Laboratory. Los Alamos National Laboratory is operated by Triad National Security, LLC, for the National Nuclear Security Administration of the U. S. Department of Energy (Contract No. 892333218NCA000001).

\section{APPENDIX A: DERIVATION OF Eq.~(5) }
\begin{figure}[htp]
    \centering
	\includegraphics[width=85mm]{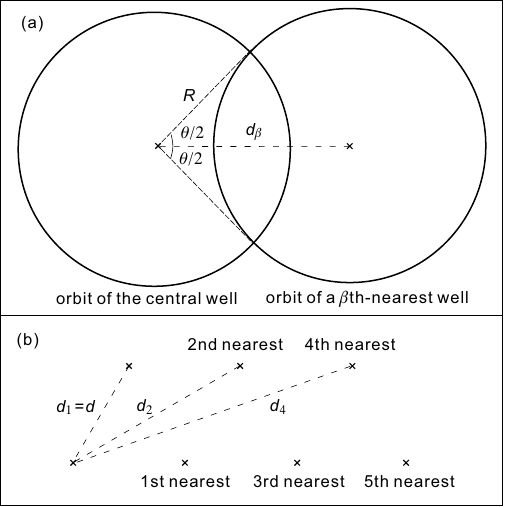}
\caption{Geometry of orbits of the centers of a potential well and one of its $\beta$th-nearest neighboring potential wells in the reference frame of moving particle (a). The arrangement of one lattice point and its $\beta$th-nearest neighboring lattice points in the 2D triangular lattice (b).}
\label{figure6}
\end{figure}

In the reference frame of the moving particle, the orbit of the center of one potential well intersects with that of its $\beta$th-nearest neighboring potential well at two points, resulting in the angle $\theta$ as shown in Fig.~\ref{figure6}(a). If we suppose the distance between the centers of the central potential well and its $\beta$th-nearest neighbor as $d_\beta$, then the radius $R$ of the orbit of each potential well center can be written as 
\begin{equation}
    R=\frac{d_\beta }{2\cos(\theta/2)}.
\end{equation}
Next, we derive the relation between $d_\beta$ and $\beta$ for the range of $1 \le \beta < 6$. In Fig.~\ref{figure6}(b), we plot one lattice point and its 1st to 5th nearest neighboring points in a triangular 2D lattice with the hexagonal symmetry. For its 1st-, 3rd-, and 5th-nearest neighbors, i.e., $\beta=1$, 3, and 5, the distance between the studied point and its neighbor can be written as $d_\beta=(\beta+1)d/2$, where $d$ is the the distance between two nearest lattice points. Similarly, for $\beta=2$ and 4, the distance between the studied point and its neighbor can be written as $d_\beta =\sqrt{[(\beta+1)d/2]^2+(\sqrt{3}d/2)^2}$. Thus, we generalize the expression of the distance between a lattice point and its $\beta$th-nearest neighbor as
\begin{equation}
    d_\beta =d\sqrt{\left[ \frac{(\beta+1)}{2} \right] ^2+\frac{[3+3\times(-1)^\beta]}{8}}.
\end{equation}
Combining Eq.~(6) and Eq.~(7) together, and also substituting $R$ using $A/m\omega^2$, we finally obtain Eq.~(5).

As $R$ varies, the orbits of the center of the six $\beta$th-nearest potential wells are able to evenly divide the orbit of the center of the central potential well into 6 or 12 parts, corresponding to the hexagonal or dodecagonal symmetry, respectively, as shown in Fig.~\ref{figure4} for $\beta=1$. In Fig.~\ref{figure4}(a), the corresponding central angle is $\theta=0$, since the two neighboring orbits are always tangent to each other. In Fig.~\ref{figure4}(b), each of the neighboring orbits always shares one twelfth of the circular orbit, as a result, the corresponding central angle is $\theta = 2\pi/12 = \pi/6$. In Fig.~\ref{figure4}(c), each of the neighboring orbits shares one sixth of the circular orbit, so that $\theta = 2\pi/6 = \pi/3$. In Fig.~\ref{figure4}(d), each of the neighboring orbits shares three twelfths of the circular orbit, resulting in $\theta = 2\pi\times3/12 = \pi/2$. Note, the four values of $\theta$ corresponding to the two symmetries are independent from the value of $\beta$.

\bibliographystyle{apsrev4-1}

\end{document}